\documentclass{aastex63}
\usepackage{amsmath}
\received{XXXX}
\revised{XXXX}
\accepted{XXXX}
\submitjournal{ApJ}

\shorttitle{Tests of Kerr Hypothesis with MAXI J1803-298}
\shortauthors{Liao, Ghasemi-Nodehi, Cui* et al.}
\graphicspath{{./}{figures/}}

\begin{document}
\title{Tests of the Kerr Hypothesis with MAXI J1803-298 Using Different RELXILL\_NK Flavors}
%\title{Very Long Baseline Array Detection of Parsec-scale Radio Emission in Dual Active Galactic Nuclei}
%\title{Parsec-scale radio properties of four dual-AGN}

\correspondingauthor{Lang Cui}
\email{cuilang@xao.ac.cn}

\author{Jie Liao}
\affiliation{Xinjiang Astronomical Observatory, CAS, 150 Science-1 Street, Urumqi 830011, China}
\affiliation{College of Astronomy and Space Science, University of Chinese Academy of Sciences, No.1 Yanqihu East Road, Beijing 101408, China}

\author[0000-0001-6113-0317]{M. Ghasemi-Nodehi}
\affiliation{Xinjiang Astronomical Observatory, CAS, 150 Science-1 Street, Urumqi 830011, China}

\author[0000-0003-0721-5509]{Lang Cui}
\affiliation{Xinjiang Astronomical Observatory, CAS, 150 Science-1 Street, Urumqi 830011, China}
\affiliation{Key Laboratory of Radio Astronomy, CAS, 150 Science 1-Street, Urumqi 830011, China}
\affiliation{Xinjiang Key Laboratory of Radio Astrophysics, 150 Science 1-Street, Urumqi 830011, China}

\author[0000-0002-0786-7307]{Ashutosh Tripathi}
\affiliation{George P. and Cynthia Woods Mitchell Institute for Fundamental Physics and Astronomy,\\ Texas A\&M University, College Station, TX 77843-4242, USA}
\affiliation{Xinjiang Astronomical Observatory, CAS, 150 Science-1 Street, Urumqi 830011, China}
% \nocollaboration{2}

\author[0000-0001-7199-2906]{Yong-Feng Huang}
\affiliation{School of Astronomy and Space Science, Nanjing University, Nanjing 210023, China}
\affiliation{Key Laboratory of Modern Astronomy and Astrophysics (Nanjing University), Ministry of Education, Nanjing 210023, China}

\author{Xiang Liu}
\affiliation{Xinjiang Astronomical Observatory, CAS, 150 Science-1 Street, Urumqi 830011, China}
\affiliation{Key Laboratory of Radio Astronomy, CAS, 150 Science 1-Street, Urumqi 830011, China}
\affiliation{Xinjiang Key Laboratory of Radio Astrophysics, 150 Science 1-Street, Urumqi 830011, China}



\begin{abstract}

Iron line spectroscopy has been one of the leading methods not only for measuring the spins of accreting black holes but also for testing fundamental physics. Basing on such a method, we present an analysis of a dataset observed simultaneously by NuSTAR and NICER for the black hole binary candidate MAXI J1803-298, which shows prominent relativistic reflection features. Various \texttt{relxill\_nk} flavors are utilized to test the Kerr black hole hypothesis. The results obtained from our analysis provide stringent constraints on Johannsen deformation parameter $\alpha_{13}$ with the highest precise to date, namely $\alpha_{13}=0.023^{+0.071}_{-0.038}$ from \texttt{relxillD\_nk} and $\alpha_{13}=0.006^{+0.045}_{-0.022}$ from \texttt{relxillion\_nk} respectively in 3-$\sigma$ credible lever, where Johannsen metric reduces to Kerr metric when $\alpha_{13}$ vanishes. Furthermore, we investigate the best model-fit results using Akaike Information Criterion and assess its systematic uncertainties.

\end{abstract}


\section{Introduction} \label{sec:introduction}

Since Einstein proposed General Relativity in late 1915, it has found applications across various physical phenomena and has undergone numerous tests in the weak field regime \citep{will2014confrontation}. Over the decades, advancements in instruments and technology have made the testing of general relativity in strong gravitational regimes a prominent and contemporary research focus. Astrophysical black holes, which can be described by the Kerr solution \citep{kerr1963gravitational, carter1971axisymmetric, robinson1975uniqueness}, serve as an ideal laboratory for probing strong gravity.

The presence of an accretion disk, nearby stars, or a potential non-vanishing electric charge of the black hole is typically negligible in the strong gravitational field near the event horizon \citep{bambi2009black, bambi2018astrophysical, cardoso2019testing}. Conversely, certain plausible macroscopic deviations from the Kerr metric arise in the presence of quantum gravity effects \citep{dvali2013black, giddings2017astronomical, giddings2018event}, exotic matter \citep{giddings2018event, herdeiro2016kerr}, and various modified theories of gravity \citep{kleihaus2011rotating, ayzenberg2014slowly, sotiriou2014black}. Consequently, testing the Kerr metric proves to be an effective approach for exploring the strong gravitational field regime.

Numerous methods for testing the Kerr hypothesis have been explored, primarily encompassing electromagnetic techniques \citep{johannsen2016sgr, Bambi:2016sac} and, in recent years, gravitational wave approaches \citep{glampedakis2006mapping, yunes2013gravitational, scientific2016tests, yunes2016theoretical}. The X-ray reflection spectrum is generally used to study the relativistic effects on the inner part of the accretion disk around BHs and to understand the properties of spacetime \citep{bambi2021towards}. There are many observational constraints already published using X-ray reflection spectrum originated from the accretion disk around Black Holes to test the Kerr hypothesis \citep{cao2018testing,xu2018study,tripathi2019constraints,tripathi2019toward,tripathi2019constraining,abdikamalov2019testing}. Among these approaches, the disk-corona model is regarded as a phenomenological model describing the relationship between the accretion disk and the ionized corona in a black hole system. 

The thermal photons emitted from the disk undergo inverse Compton scattering within the corona, characterized by high temperatures (approximately 100 \text{keV}), and some of them are reflected to the disk, which is the so-called reflection spectrum. The most prominent features of the reflection spectrum are often characterized by the iron $K\alpha$ line around 6.4 keV, depending on the ionization of iron atoms, and the Compton hump peaked around \text{20--30} \text{keV}. In the rest frame of the gas, fluorescent emission lines display narrow profiles; however, in a strong gravity region, the $K\alpha$ line undergoes broadening due to relativistic effects, which indicates that it is one of the most potential tools that can be used to test relativistic effects in strong gravity region \citep{cao2018testing,abdikamalov2019testing,tripathi2020testing}.

The \texttt{relxill\_nk} model\footnote{\url{https://github.com/ABHModels/relxill_nk}}\citep{bambi2017testing,abdikamalov2019public}, an extension of the \texttt{relxill} package\footnote{\url{http://www.sternwarte.uni-erlangen.de/~dauser/research/relxill/}}\citep{dauser2013irradiation,garcia2014improved}, is a versatile tool for analyzing the reflection features of a geometrically thin and optically thick disk in non-Kerr spacetimes. The model employs a parametric black hole spacetime metric, where a set of deformation parameters parameterizes deviations from the Kerr solution. The Kerr metric is recovered when all the deformation parameters vanish \citep{cao2018testing,abdikamalov2019testing,tripathi2020testing}.

This paper details the spectral analysis conducted on the observational data during the outburst of the Galactic black hole binary candidate MAXI J1803-298. The outburst of this black hole binary candidate was first captured by the Gas Slit Camera of the Monitor of All-sky X-ray Image (MAXI/GSC) nova alert system at 19:50 UT on May 1st, 2021, located at R.A.=$270.923^\circ$, Dec=$-29.804^\circ$ (J2000) \citep{serino2021maxi}. A comprehensive multiwavelength follow-up of the discovery outburst and the timing analysis of the black hole candidate MAXI J1803$-$298 is presented in \citet{sanchez2022hard} and \citet{zhu2023timing}. These findings indicate a state transition from the low/hard state to the hard intermediate state, followed by the soft intermediate state, and ultimately reaching the high/soft state. The works of \citet{feng2022spin} and \citet{coughenour2023reflection} previously examined variability and reflection features, revealing a notable relativistically broadened iron line component in the spectrum with an extraordinarily high value of spin parameter.

Based on the reflection feature from MAXI J1803-298 within a soft intermediate state, we mainly test the Kerr metric with different flavors of model \texttt{relxill\_nk} and then discuss the impact of different \texttt{relxill\_nk} flavors in fitting data from NuSTAR and NICER in this work. Stringent constraints on its parameters have been obtained, and these findings reveal no significant distinctions among different flavors of \texttt{relxill\_nk}. Subsequently, the Akaike Information Criterion (AIC) was employed to assess the congruence among diverse models for the purpose of selecting the best-fitting model.

The contents of this paper are organized as follows. In 
Section \ref{sec:Observation and Data Reduction}, we present the observations of MAXI J1803-298 and data reduction from NuSTAR and NICER. In Section \ref{sec:Spectrum Analysis}, we present our spectrum analysis with different \texttt{relxill} flavors and \texttt{relxill\_nk} flavors. In Section \ref{sec:Discussion and Conclusions}, we discuss our results, estimate the systematic uncertainties, and give conclusions in the end.

% \url{http://journals.aas.org/authors/aastex.html}.

\section{Observation and Data Reduction} 
\label{sec:Observation and Data Reduction}

\begin{deluxetable*}{cccc}
\tablenum{1}
\tablecaption{Observations Analyzed in the Present Work\label{tab:Observation}}
\tablewidth{0pt}
\tablehead{
\colhead{Mission} & \colhead{Observation ID} & \colhead{Start Date} & \colhead{Exposure (s)}
}
% \decimalcolnumbers
\startdata
% NuSTAR & 90702316002 & 2021-05-05 & 26554 \\
% NuSTAR & 80701332002 & 2021-05-14 & 31603 \\
NuSTAR & 90702318002 & 2021-05-23 & 12922 \\  
% NuSTAR & 90702318003 & 2021-06-17 & 15584 \\
NICER & 4202130110 & 2021-05-23 & 7508 \\
\enddata
% \tablecomments{Exposure time is the dead-time-corrected on-source live time for one NuSTAR module, FPMA.}
\end{deluxetable*}

\subsection{Observations} \label{subsec:obsevations}
 MAXI J1803-298 was observed by various X-ray missions, including simultaneous observations by NuSTAR and NICER on May 23, 2021. Observation IDs and their exposure times are reported in Table \ref{tab:Observation}. The focus of this work is the observation with ObsID 90702318002 from NuSTAR/(\text{FPMA}$|$\text{FPMB}) and ObsID 4202130110 from NICER/XTI followed by works in \citet{feng2022spin}. Assuming that the Kerr solution describes the spacetime metric around the black hole, \citet{feng2022spin} and \citet{coughenour2023reflection} estimated the black hole's spin and well-constrained features of the reflection spectrum. We followed their works but mainly employed the \texttt{relxill\_nk} with different flavors to estimate the deformation parameter $\alpha_{13}$ of the Johannsen metric. The line element of the Johannsen spacetime is reported in the Appendix, where we also list the main properties of this black hole metric.

\begin{figure}[ht!]
    \centering
    \includegraphics[width=0.5\linewidth, height=0.25\linewidth]{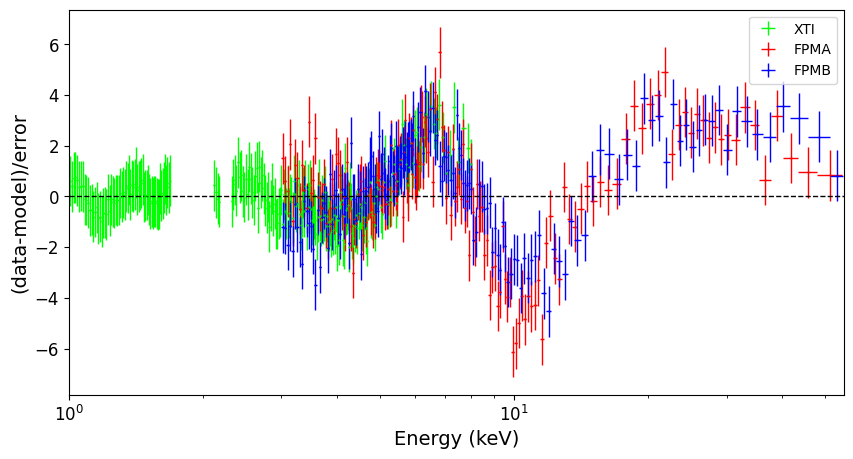} % 设置图片宽度为页面宽度的80%，高度为页面高度的60%
    \caption{Normalized residuals for an absorbed power-law spectrum + disk blackbody spectrum [in XSPEC language, \texttt{tbabs$\times$(diskbb+powerlaw)}]. Green crosses are for XTI/NICER data, red crosses are for FPMA/NuSTAR data, and blue crosses are for FPMB/NuSTAR data. Data have been rebinned for visual clarity. \label{fig:iron}}
\end{figure}

\subsection{Data reduction} \label{subsec:reduction}
The raw data obtained from the NuSTAR detectors, denoted as FPMA and FPMB, are processed by the standard pipeline\footnote{\url{https://heasarc.gsfc.nasa.gov/docs/nustar/analysis/}} \texttt{NUPIPELINE 0.4.9} in \texttt{HEASOFT v6.32} using the latest calibration database \texttt{CALDB v20230816}.
 The Good time interval (GTI) files are produced using \texttt{nuscreen}, and we remove some obvious variabilities in the NuSTAR light curve by tool \texttt{fv} to preclude the impact of flare and dip events on the spectrum analysis.
 %\textcolor{red}{This should be discussed in detail. What exactly you cleared from the light curve and how you cleared it in your spectrum.} 
 The source events are extracted from a circular region centered on MAXI J1803-298 with a radius of $120''$. The background region is a circle of the same size taken far from the source region to avoid any contribution from the source. We use the \texttt{nuproducts} to generate the spectra and other products. The FPMA and FPMB spectra are grouped to have a minimum count of 25 photons per bin for the chi-squared statistics to be applicable. The NuSTAR data are modeled over the 3–55 keV band for further analysis.

Based on the latest calibration file, the NICER data are processed following the standard steps\footnote{\url{https://heasarc.gsfc.nasa.gov/docs/nicer/analysis_threads/}}. We use the tool \texttt{nicer12} and \texttt{nibackgen3C50} to extract the source and background spectra, respectively. The response matrix and ancillary file are generated by \texttt{nicerrmf} and \texttt{nicerarf}. In this work,  the spectrum from NICER's data is fitted over the energy band of 1.0-8.0 keV, ignoring the energy range of 1.7-2.1 keV (calibration residuals remain found in Si band) and 2.2-2.3 keV (calibration residuals remain in Au edges), respectively. We also group the XTI spectra to have a minimum count of 25 photons per bin.

\section{Spectrum Analysis}\label{sec:Spectrum Analysis}
Spectra are modeled using XSPEC v12.13.1 with $\chi^2$ statistics. To fit these spectra simultaneously from FPMA, FPMB, and XTI, a constant multiplicative factor was included in each model, set to 1.0 for XTI, and allowed to vary freely for FPMA and FPMB to eliminate the calibration differences in different instruments in the
process of joint fit. To begin with, We fit the NuSTAR spectrum with a galactic absorbed power-law component from the corona and a thermal spectrum from the disk. The normalized residual is shown in Figure \ref{fig:iron}. This residual presents a strong reflection feature with a broad iron line around 6.4 keV and a Compton hump with a peak at 20-30 keV, with $\chi^{2}_{\nu}=3127.37/2290$.

\begin{deluxetable*}{ccccccc}
\tablenum{2}
\tablecaption{Best-fit Values for the Kerr Model I1-I4.\label{tab:relxill_parameters}}
\tablewidth{500pt}
\tabletypesize{\scriptsize}
% \tablewidth{0pt}
\tablehead{
\colhead{} & \colhead{Parameter} & \colhead{Model I1} & \colhead{Model I2}  & \colhead{Model I2$^+$} & \colhead{Model I3} & \colhead{Model I4}
}
% \decimalcolnumbers
\startdata
{\texttt{tbabs}} & {}  & {}& {} & {}  & {} & {}\\
{$N_H(10^{21} cm^{-2}$)}  & {Hydrogen column density}& {$3.68 \pm 0.04$} & {$3.39 \pm 0.07$} & {$3.54 \pm 0.12$} & {$3.31\pm 0.04$} & {$3.16 \pm 0.06$}\\
\hline
{\texttt{diskbb}} & {} & {} & {} & {}  & {} & {}\\
{$T_{in}(keV)$} & {Temperature of disk} & {$0.813 \pm 0.017$} & {$0.768 \pm 0.007$} & {$0.822 \pm 0.009$} & {$0.765 \pm 0.005$} & {$0.793 \pm 0.010$}\\
{$N_{diskbb}$} & {Normalization} & {$425 \pm 43$} & {$612 \pm 30$} & {$433 \pm 31$} & {$598 \pm 19$} & {$546 \pm 33$}\\
\hline
{\texttt{relxill flavor}} & {}  & {\texttt{relxill}}& {\texttt{relxillCP}} & {\texttt{relxillCP}}  & {\texttt{relxilllp}} & {\texttt{relxilllpCP}}\\
{h} & {Height of the corona}  & {$-$}& {$-$} & {$-$}  & {$<11.7$} & {$<10.8$}\\
% {$\beta$(v/c)} & {Velocity of the primary source}  & {$-$}& {$-$} & {$-$}  & {$0^*$} & {$0^*$}\\
{$q_{in}$} & {Emissivity index in the inner region} & {$10.0_{-4.9}$} & {$9.64^{}_{-2.1}$} & {$10.0^{}_{-5.09}$} & {$-$} & {$-$}\\
{$q_{out}$} & {Emissivity index in the outer region} & {$5.7 \pm 5.7$} & {$6^*$} & {$6^*$}  & {$-$} & {$-$}\\
{$R_r$(M)} & {Break radius} & {$2.13 \pm 1.07$} & {$2.37 \pm 0.79$} & {$2.15 \pm 0.30$} & {$-$} & {$-$}\\
{$a^*$} & {Black hole spin} & {$0.989 \pm 0.005$} & {$0.993 \pm 0.007$}  & {$0.984 \pm 0.006$} & {$0.998^*$} & {$0.998^*$}\\
{$i$(deg)} & {Inclination angle} & {$72.5 \pm 1.5$} & {$72.5 \pm 1.8$} & {$68.2 \pm 1.7$} & {$50.3 \pm 2.2$} & {$44.1 \pm 1.9$}\\
{$R_{in}$} & {Disk inner radius} & {$-1^*$} & {$-1^*$} & {$-1^*$}  & {$-1^*$} & {$-1^*$}\\
{$\Gamma$} & {Photon Index} & {$2.34 \pm 0.03$} & {$2.25 \pm 0.02$} & {$2.22 \pm 0.05$} & {$2.23 \pm 0.02$} & {$2.17 \pm 0.04$}\\
{$log\xi$} & {Ionization state of disk} & {$3.78 \pm 0.28$} & {$4.44 \pm 0.21$} & {$3.35 \pm 0.09$} & {$4.70 \pm 0.19$} & {$3.97 \pm 0.14$}\\
{$logN$} & {The density of the accretion disk} & {$15^*$} & {$15^*$} & {$18.3 \pm 0.5$}  & {$15^*$} & {$18.0 \pm 0.6$}\\
{$Fe$} & {Iron abundance} & {$1.17 \pm 0.42$} & {$5.63 \pm 2.90$} & {$1.30 \pm 0.28$}  & {$7.07 \pm 2.19$} & {$4.35 \pm 1.60$} \\
{$E_{cut}(kT_e)(keV)$} & {Energy cut-off (or $T_{corona}$)} & {$300^*$} & {$100^*$} & {$100^*$}  & {$300^*$} & {$100^*$}\\
{$R_f$} & {Reflection fraction} &{$1.92 \pm 0.87$} & {$1.77 \pm 0.52$} & {$1.74 \pm 0.47$} & {$1.16 \pm 0.42$} & {$0.66 \pm 0.38$}\\
{$norm(10^{-2})$} & {Normalization} & {$1.9 \pm 0.4$} & {$1.5 \pm 0.2$} & {$1.2 \pm 0.3$} & {$2.0 \pm 0.7$} & {$2.2 \pm 1.2$}\\
\hline
{\texttt{gaussian}} & {} & {} & {} & {}  & {} & {}\\
{$E_{line}$} & {Absorption line energy in keV} & {$7.12 \pm 0.06$} & {$7.11 \pm 0.07$} & {$7.11 \pm 0.06$} & {$7.33 \pm 0.10$} & {$7.06 \pm 0.10$}\\
{$\sigma_{E}$} & {Line width in keV} & {$0.08 \pm 0.09$} & {$0.06 \pm 0.10$} & {$0.08 \pm 0.08$} & {$0.51 \pm 0.11$} & {$0.78 \pm 0.07$}\\
\hline
{$C_{FPMA}$} & {Cross-normalization} & {$1.023$} & {$1.023$} & {$1.023$} & {$1.023$} & {$1.024$}\\
{$C_{FPMB}$} & {Cross-normalization} & {$0.996$} & {$0.997$} & {$0.997$} & {$0.997$} & {$0.997$}\\
\hline
{\texttt{$\chi^{2}/\nu$(reduced)}} & {$-$} & {$2520/2279$} & {$2545/2280$} & {$2492/2279$} & {$2509/2282$} & {$2507/2281$}\\
 {} & {} & {$=1.106$} & {$=1.116$} & {$=1.093$} & {$=1.099$} & {$=1.099$}\\
\enddata
\tablecomments{Best-fit values of initial Model I1-I4:  $\texttt{tbabs}\times(\texttt{diskbb}+{\texttt{relxill(flavor)}}+\texttt{gaussian})$, in Xspec language (\texttt{relxill}, \texttt{relxillCP}, \texttt{relxillp} and \texttt{relxillpCP}, respectively), with errors calculated within a 90\% confidence interval by $\chi^2$ statistic. $*$ indicates that the parameter is frozen in the fit. $^+$ indicates that the parameter $logN$ is free. $R_{in}=-1$ means that $R_{in}$ is set at the ISCO radius. The radial coordinate of the outer edge of the accretion disk is fixed at 400. $i$ is allowed to vary from $3^\circ$ to $80^\circ$. $a^*$ is allowed to vary from $-0.998$ to $0.998$. $logN$ in \text{relxillCP} flavors are allowed to range from 15 to 19. When the lower/upper uncertainty is not reported, the 90\% confidence level reaches the boundary (or the best-fit is at the boundary).}
\end{deluxetable*}

To study the strong relativistic reflection composition in this source, our primary full XSPEC model involves substituting the power-law component with standard reflection models, denoted as {\texttt{relxill(flavor)}}, which encompasses {\texttt{relxill}}, {\texttt{relxillCP}}, {\texttt{relxillp}}, and {\texttt{relxillpCP}}. The initial model is described by $\texttt{constant}\times\texttt{tbabs}\times(\texttt{diskbb}+{\texttt{relxill(flavor)}})$, where the specific flavors are respectively marked as I1-I4. \texttt{tbabs} describes the galactic absorption and has only one parameter; column density ($N_H$) along the line of sight. $N_H$ is set to be free while fitting the spectra. \texttt{diskbb} describes the thermal spectrum from the accretion disk. \texttt{relxill(flavor)} component describes the power-law and reflection composition. As it is widely believed that the accretion disk approaches the ISCO between the intermediate and soft states, we assume that the inner edge of the accretion disk is at the ISCO (equal to $6GM/c^2=6R_g$ for a Schwarzchild BH, or just 1 $R_g$ for a maximally spinning Kerr BH), and the outer radius is set to 400 $R_g$ where $R_g$ is the gravitational radius. The emissivity profile in model I1-I2 is modeled with a broken power-law emissivity profile with three parameters, and let them to free. Due to the limited fitting energy range, we additionally set the electron temperature of the corona (kTe) to 1/3 of the default value of the power law cutoff energy ($E_{cut}$ = 300 keV), i.e., kTe = 100 keV. Because inclination is measured by a high value from the same and another NuSTAR's observation from \citet{feng2022spin} and \citet{coughenour2023reflection}, we also left it free to vary in the same way. In the lamppost model I3 and I4, we freeze the spin parameter to 0.998. Besides, a Fe XXVI absorption line at 7.0 keV has been detected during its intermediate state in \citet{zhang2024diskwinds}, we also added a \texttt{gaussian} component in our model I1-I4.

After fitting the data with the model, we found the "standard" weighting scheme in XSPEC often resulted in overfitting the data, yielding a reduced $\chi^2 < 1$.  This is a known issue discussed in \citet{galloway2020multi} regarding handling the low-count bins in \texttt{XSPEC v.12}. The Churazov weighting scheme is an alternative to the standard weighting used for fitting data \citep{churazov1996mapping}. It adjusts the weight by taking into account the counts in neighboring channels. This ensures that local extrema are not given disproportionate weight, leading to a smoother overall weighting. The result is a more accurate and reliable data fitting. Our best-fit results of Model I1-I4 are shown in Table \ref{tab:relxill_parameters}, and all errors are reported at the 90\% credible level.

\begin{figure}[ht!]
  \centering
  \begin{minipage}{0.49\linewidth}
    \centering
    \includegraphics[width=\linewidth]{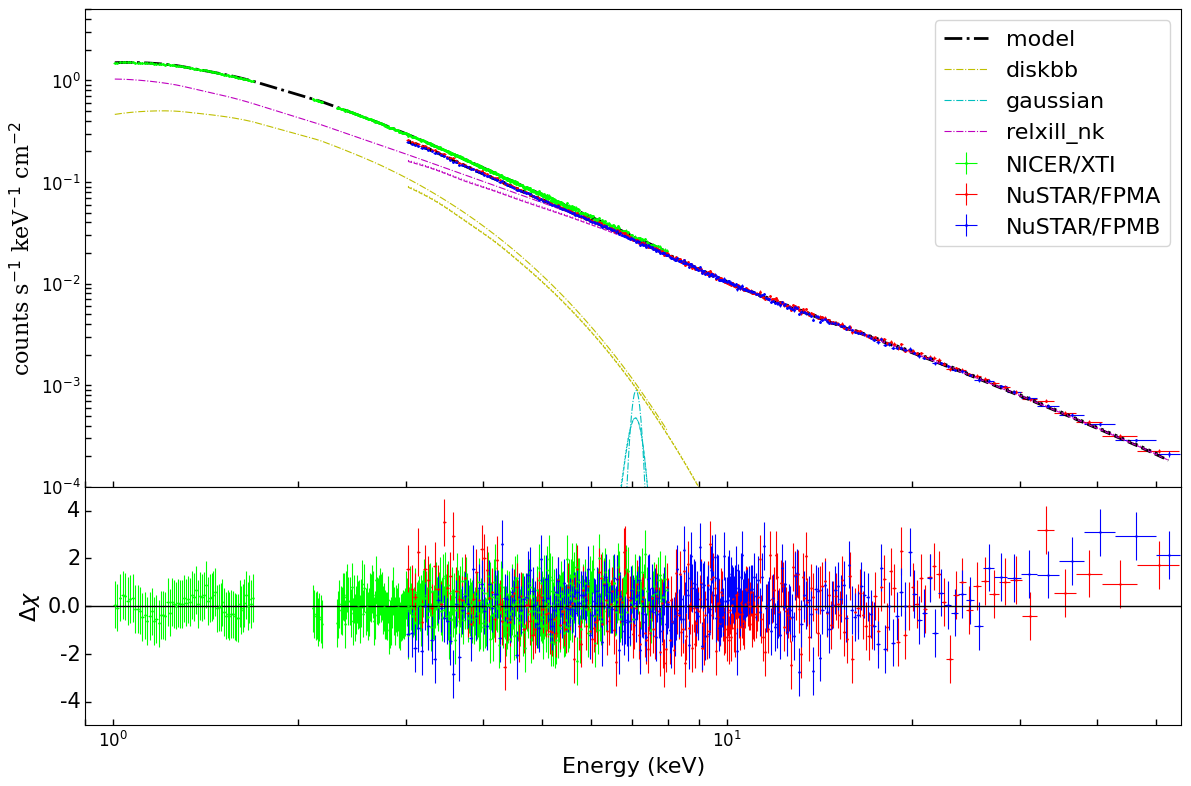}
  \end{minipage}
  \hfill
  \begin{minipage}{0.49\linewidth}
    \centering
    \includegraphics[width=\linewidth]{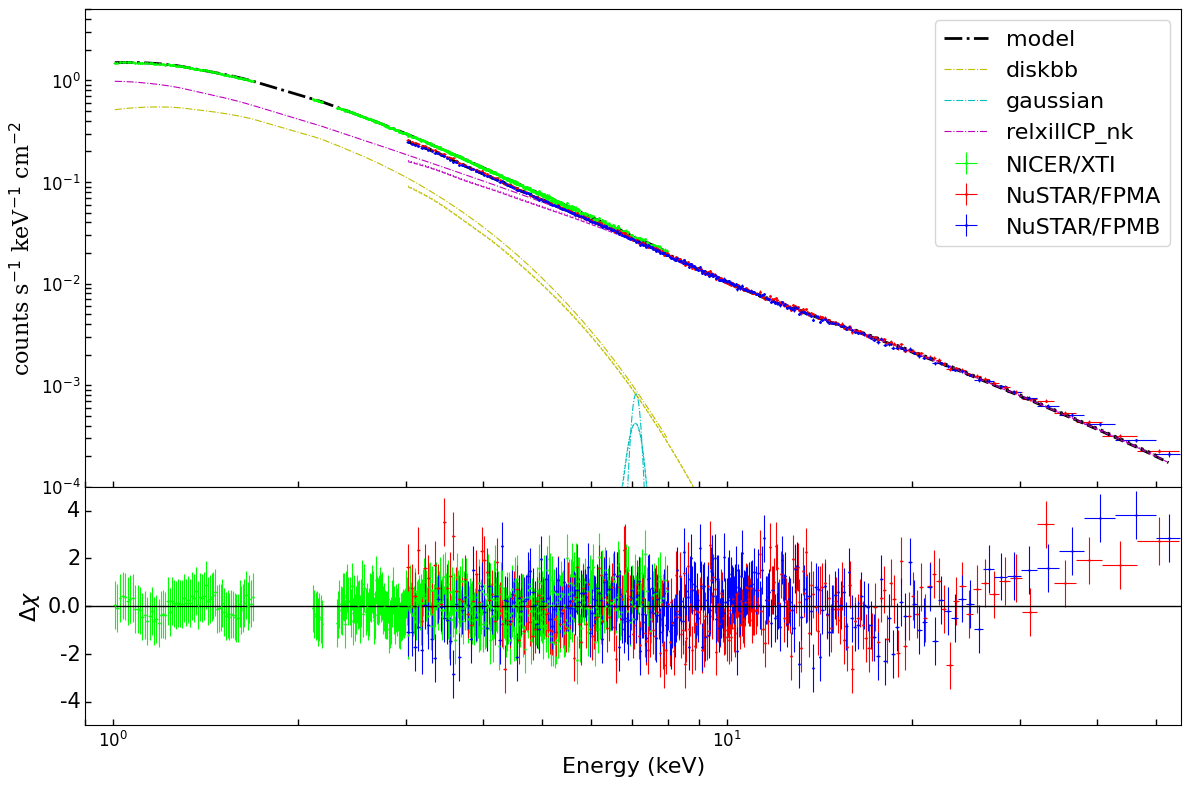}
  \end{minipage}
  
  \begin{minipage}{0.49\linewidth}
    \centering
    \includegraphics[width=\linewidth]{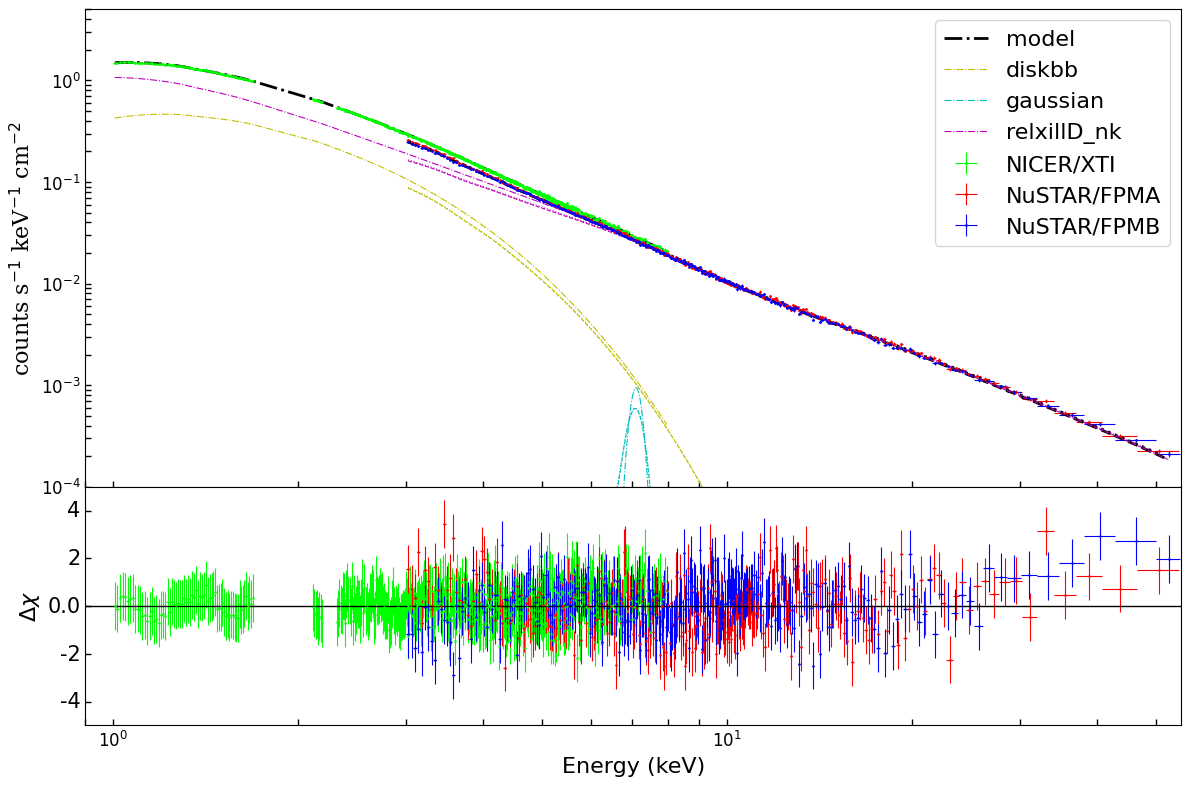}
  \end{minipage}
  \hfill
  \begin{minipage}{0.49\linewidth}
    \centering
    \includegraphics[width=\linewidth]{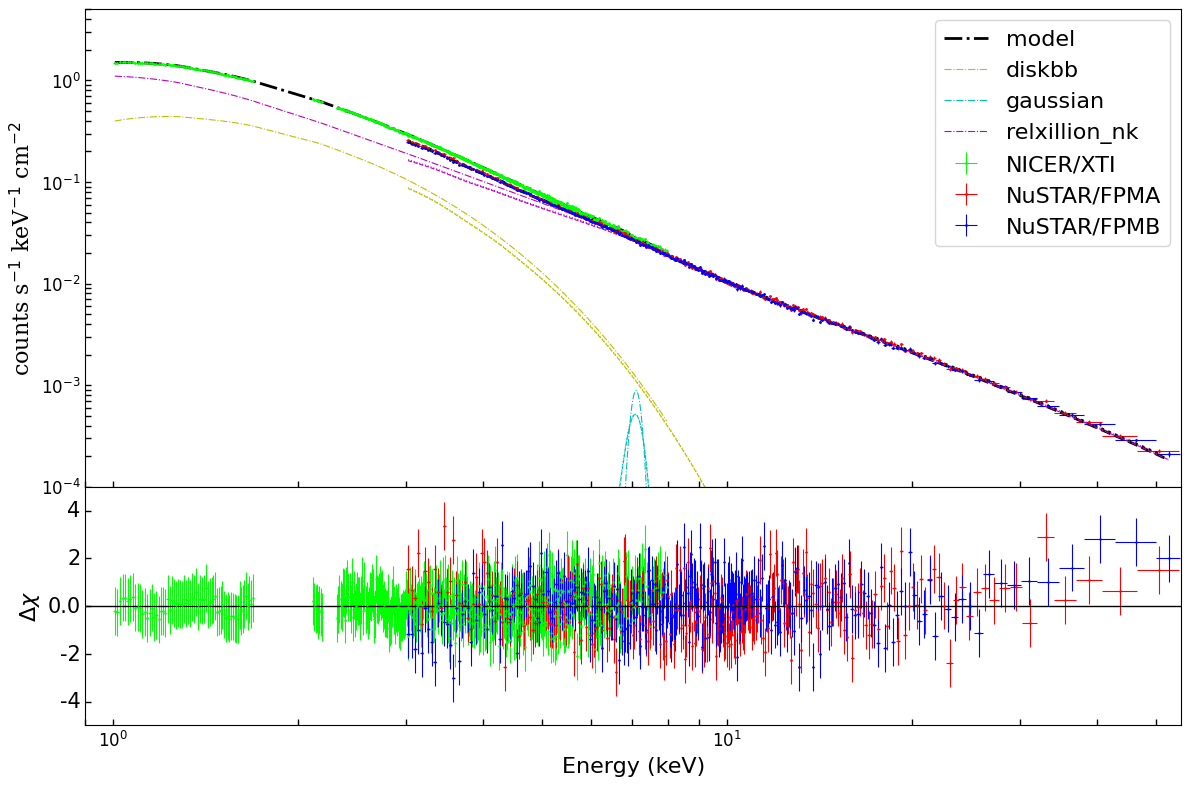}
  \end{minipage}
  
  \caption{The best-fit results of unfolded spectra and normalized residuals of Model B1-B4. Upper quadrants show total models (black), \texttt{relxill\_nk} (magenta), and \text{diskbb} (yellow). For each subgraph, dash-dot, dotted, and dashed lines present the same model for NICER/XTI, NuSTAR/FPMA, and NuSTAR/FPMB data, respectively. Lower quadrants show normalized residuals. NICER/XTI, NuSTAR/FPMA, and NuSTAR/FPMB data are plotted in green, red, and blue, respectively. Data have been rebinned for visual clarity.}
  \label{fig:model}
\end{figure}

We then substitute the \texttt{relxill(flavors)} model with various \texttt{relxill\_nk} flavors: the \texttt{relxill\_nk} (default model), \texttt{relxillCP\_nk} (\text{nthcomp} Comptonization for the coronal spectrum), \texttt{relxillD\_nk} (variable disk electron density), and \texttt{relxillion\_nk} (the value of the ionization parameter varies as the radial coordinate r increases). We conduct joint fitting on the data. The models in XSPEC language are \texttt{constant$\times$tbabs$\times$(diskbb+relxill\_nk(flavor)+gaussian)}, which are respectively marked as A1-A4 (set $\alpha_{13}$ to 0) and B1-B4 (set $\alpha_{13}$ to free). More details of the extension of the \texttt{relxill\_nk} model can be found in \citet{abdikamalov2019public}. We apply the same bounds to the emissivity profile and the break radius as we did for Model I1-I2.

Within the framework of Johannsen spacetime, it is assumed that, with the exception of $\alpha_{13}$, the remaining three deformation parameters, namely $\alpha_{22}$, $\epsilon_3$ and $\alpha_{52}$, are held constant at zero for simplicity in this work.
Here, a non-zero former signifies the departure from the Kerr metric. After we free $\alpha_{13}$, the best-fit results from our model B1-B4 are shown in Table \ref{tab:Parameters}, and its unfolded spectra and normalized residuals are shown in Figure \ref{fig:model}.

\begin{figure}[ht!]
    \centering
    \includegraphics[width=1.0\linewidth, height=1.0\linewidth]{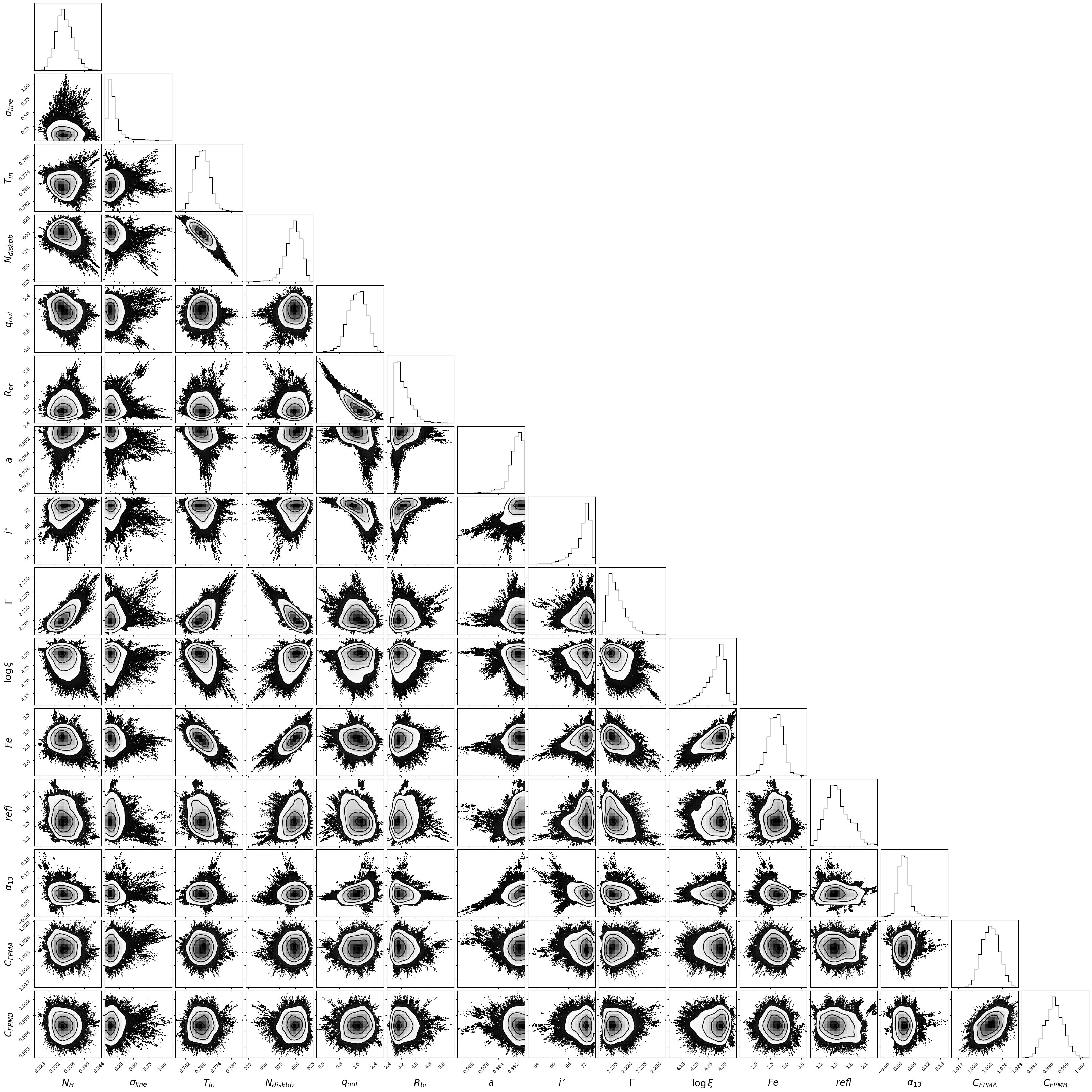} 
    \caption{The corner plot for free parameters in Model B1 after MCMC runs is used in this work. The 2D plots report the $1$, $2$, and $3\sigma$ confidence contours. \label{fig:relxill_nk}}
\end{figure}

\begin{figure}[ht!]
    \centering
    \includegraphics[width=1.0\linewidth, height=1.0\linewidth]{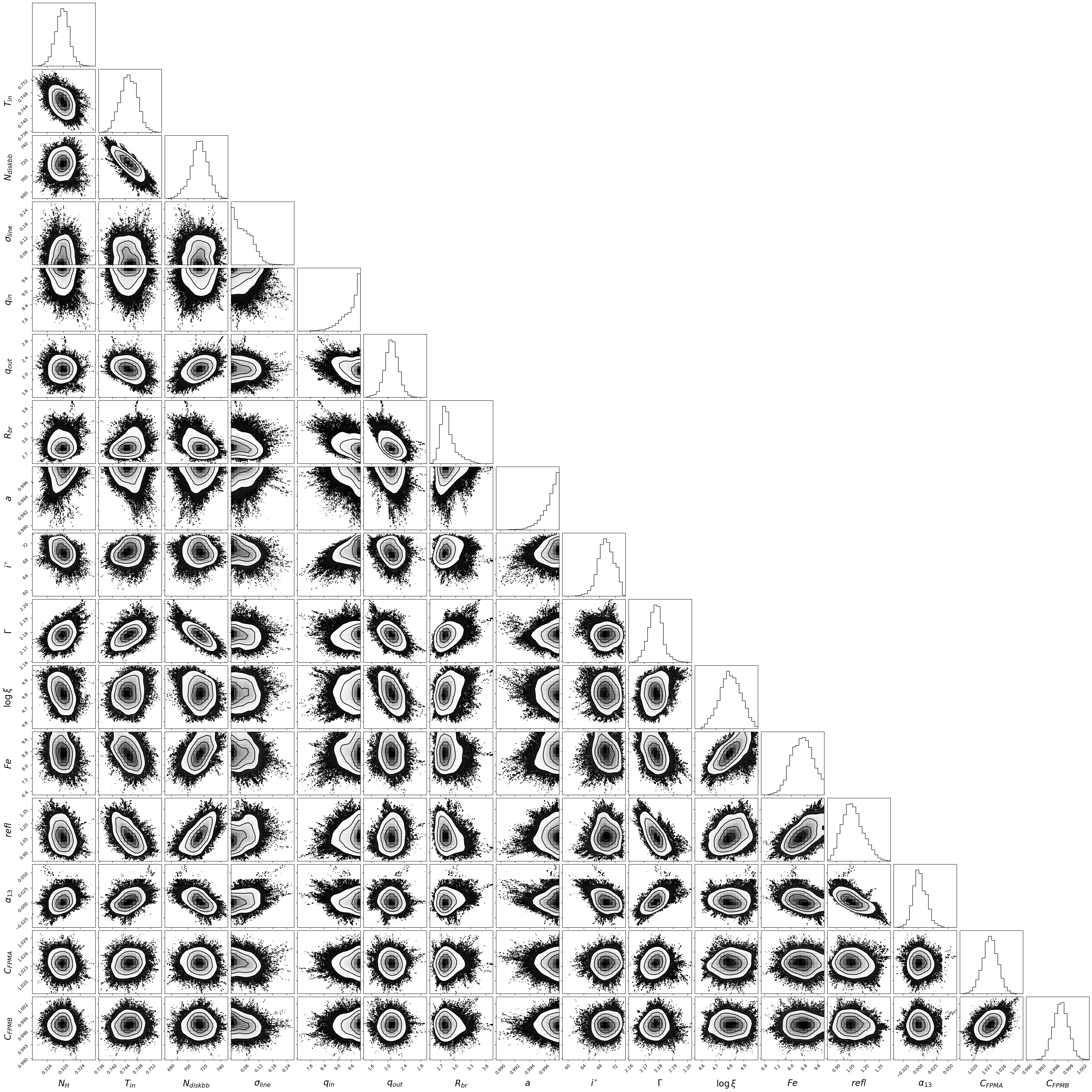} 
    \caption{The corner plot for free parameters in Model B2 after MCMC runs is used in this work. The 2D plots report the $1$, $2$, and $3\sigma$ confidence contours. \label{fig:relxillCP_nk}}
\end{figure}

\begin{figure}[ht!]
    \centering
    \includegraphics[width=1.0\linewidth, height=1.0\linewidth]{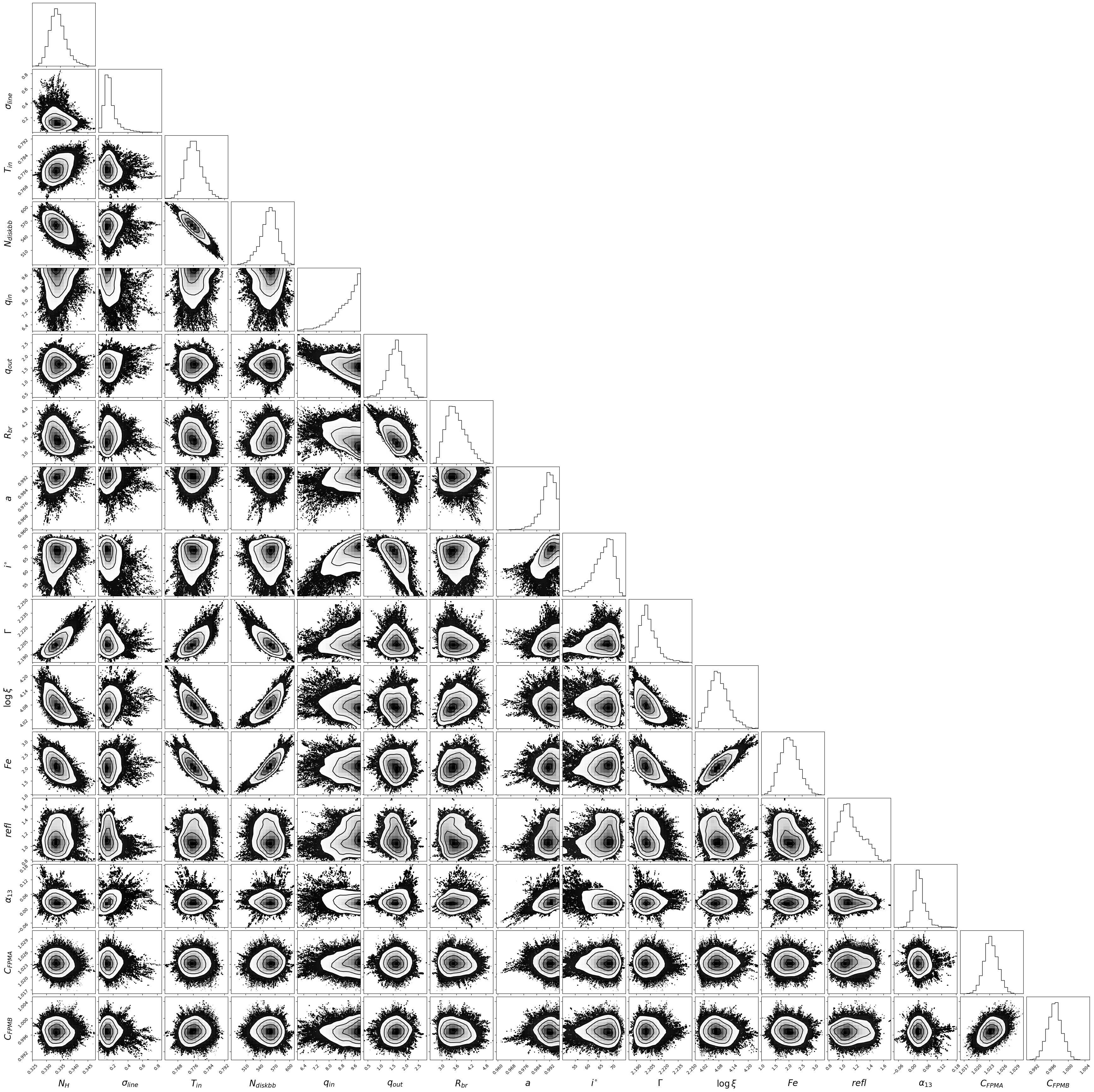} 
    \caption{The corner plot for free parameters in Model B3 after MCMC runs is used in this work. The 2D plots report the $1$, $2$, and $3\sigma$ confidence contours. \label{fig:relxillD_nk}}
\end{figure}

\begin{figure}[ht!]
    \centering
    \includegraphics[width=1.0\linewidth, height=1.0\linewidth]{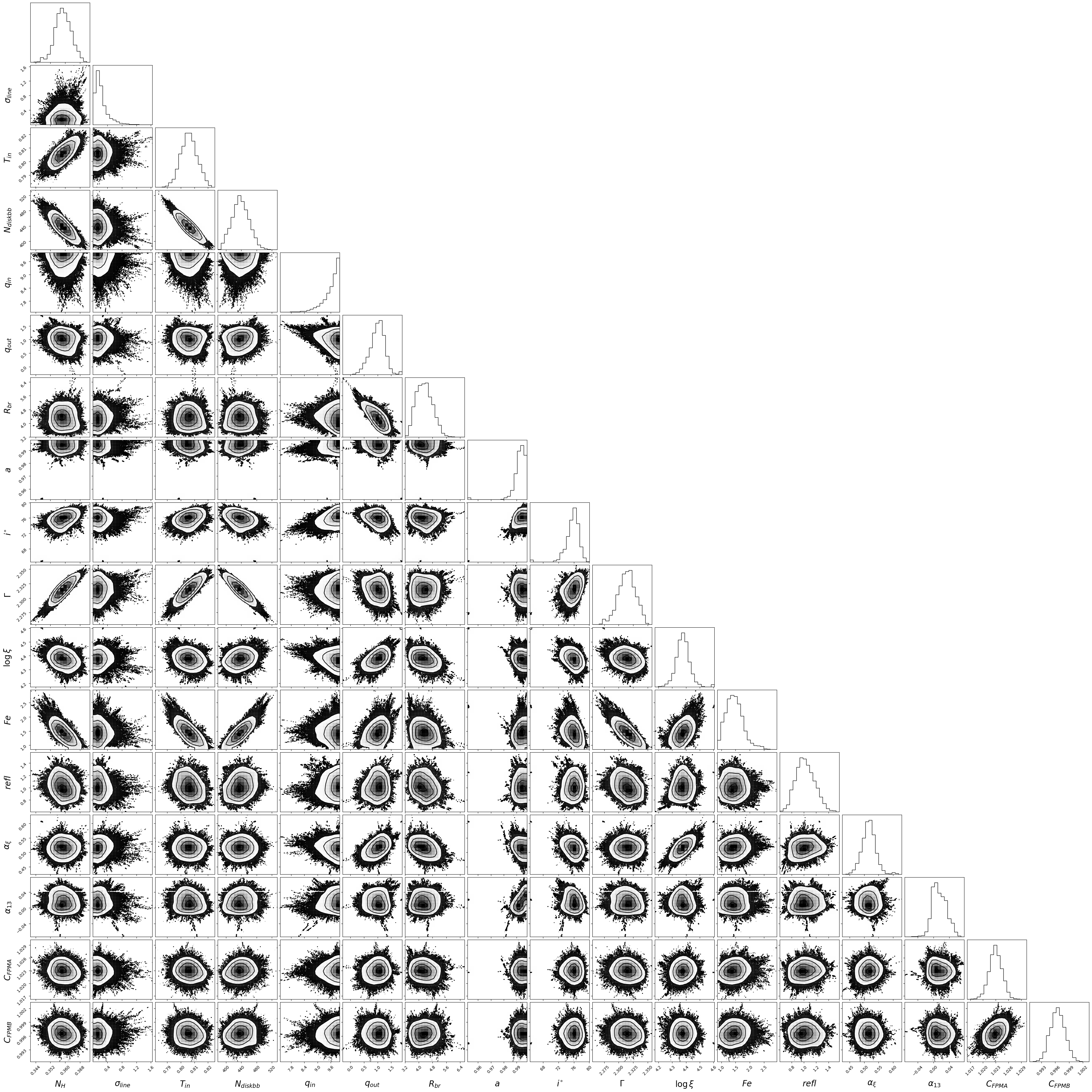} 
    \caption{The corner plot for free parameters in Model B4 after MCMC runs is used in this work. The 2D plots report the $1$, $2$, and $3\sigma$ confidence contours. \label{fig:relxillion_nk}}
\end{figure}

 The chi-square statistics is employed to determine the best-fit parameters, serving as the prior distribution for Markov Chain Monte Carlo (MCMC) analysis in the fitting process. The MCMC samples were generated using the Goodman \& Weare algorithm embedded in the XSPEC software\footnote{\url{https://github.com/zoghbi-a/xspec_emcee}}. The chains were run with 40 walkers, each comprising 25000 iterations, with an initial burn-in phase of 1000 steps. Thus, a total of $1 \times 10^6$ samples were obtained. The 90\% confidence intervals, indicative of purely statistical uncertainties across the entire parameter chain, for the free parameters within the best-fit results of models B1-B4 derived from the MCMC simulations, are detailed in Table \ref{tab:Parameters}.  Furthermore, Figure \ref{fig:relxill_nk}, \ref{fig:relxillCP_nk}, \ref{fig:relxillD_nk}, and \ref{fig:relxillion_nk} showcase corner plots illustrating the one- and two-dimensional projections of the posterior probability distributions for the relevant free parameters corresponding to Model B1-B4, respectively, as a result of the MCMC analysis.

 % \textcolor{green!70!black}{Table \ref{tab:relxill_parameters}}

\section{Discussion and Conclusions} 
\label{sec:Discussion and Conclusions}
After scrutinizing the spectra of MAXI J1803-298 from NuSTAR and NICER utilizing state-of-art relativistic reflection models in the preceding section, we derived precise and accurate constraints within these models employed in the present work. Our findings and conclusions are presented below.

In this work, we mainly used the \texttt{relxill} flavors and \texttt{relxill\_nk} flavors to fit the data. Our best-fit values of our employed models I1-I4 and A1-B6 are shown in Table \ref{tab:relxill_parameters} and Table \ref{tab:Parameters}, respectively. It is essential to recognize that direct comparisons of the $\chi^2$ values between Models I1-I4 utilizing \texttt{relxill} flavors and Models A1-B6 employing \texttt{relxill\_nk} flavors cannot definitively ascertain the superior model due to variations in the calculation of the transfer function. As such, the minor discrepancies observed in the $\chi^2$ values between these two sets of models will not impact the subsequent discussion.

\subsection{Fitting with \texttt{Relxill} Flavors}\label{sec:Relxill flavors}        
We first compare our best-fit results of model A with results shown on \citet{feng2022spin}. Our findings showcase a relatively lower galactic absorption coefficient$\sim 3.0 \times 10^{21} cm^{-1}$, which aligns with results in \citet{Homan2021ATel}. However, the best-fit values in Table \ref{tab:relxill_parameters} show a slightly higher galactic absorption coefficient than the results in \citet{chandApJHIS}, which may be caused by the outflowing disk winds and the moving clouds during the outburst. 

In the initial model I1-I2, the results obtained with \texttt{relxill} and \texttt{relxillCP}, which are standard models for relativistic reflection within the Kerr paradigm, incorporate emissivity that can be parameterized either through an empirically defined broken power law. 
Our results shown in Table \ref{tab:relxill_parameters}, indicative of a black hole with extremely high spin and a highly inclined accretion disk  (approximately $70^\circ$). Particularly, there shows $\Delta\chi^2>45$ when the density of the accretion disk is set free, denoted to Model I2$^+$. This suggests a high ionization density disk near $\sim 10^{18}$ cm$^{-3}$, within MAXI J1803-298, which aligns with \citet{coughenour2023reflection}. Thus, their fit results with a higher galactic absorption coefficient $\sim 5.0 \times 10^{21}$ cm$^{-1}$ with only a low ionization density disk model \texttt{relxillCP} in \citet{feng2022spin} appear to be inconclusive.

We subsequently incorporate lamppost models I3 and I4, which posit a lamppost source located along the rotational axis, to evaluate the proximity of the corona to the disk, setting the spin parameter to 0.998. The results indicate that the height of the compact corona above the disk is constrained to $h<10.8 R_g$, suggesting a still pronounced reflection component in the data observed during its SIMS.

\subsection{Fitting with \texttt{Relxill\_nk} Flavors}\label{sec:Relxill_nk flavors}
Afterward, we transition to the \texttt{relxill\_nk} flavors to investigate the correlation between deformation parameters $\alpha_{13}$ and the spin parameter $a^*$ based on its reflection feature. As is shown in model B1-B4 in Table \ref{tab:Parameters}, it is evident that when modeling the emissivity profile with a broken power law using the \texttt{relxill\_nk} flavors, consistently high spin values ($a^* > 0.98$) are observed. Additionally, there is a notable pattern of high values for the emissivity index $q_{in}$ and low values for $q_{out}$. This further supports our fit results of a compact corona near the black hole modeled by Model I3-I4 in Section \ref{sec:Relxill flavors}.

In model B2 with \texttt{relxillCP\_nk}, in which the parameter $kT_e$ represents the temperature of the corona, we found a stringent constraint on the deformation parameter assuming the Johannsen spacetime \citep{johannsen2013regular}. However, this constraint on $\alpha_{13}$ should be excluded due to its bad parameter estimation, details are presented in Section \ref{the best-fit model}. In addition, since the electron density is fixed to a low value of $10^{15}$ cm$^{-3}$ rather than a high disk density in the disk in Model B1 and Model B2. We employed \texttt{relxillD\_nk} for data fitting, where the parameter linked to the disk electron density, $logN$, which was frozen to 18 due to the high disk density we got from Model I2$^+$ in section \ref{sec:Relxill flavors}. Notably, when the high disk density is set to 18 in Model B3, it shows $\Delta\chi^2>14$, but it will not significantly change the constraint on $\alpha_{13}$.
The contours of spin versus deformation parameter $\alpha_{13}$ are shown in Figure \ref{fig:a13_a}.

In previous \texttt{relxill\_nk} flavors, the ionization parameter $\xi$ was not considered as a variable. To address this, we also applied the \texttt{relxillion\_nk} model \citep{abdikamalov2021implementation}, utilizing \texttt{xillver} table for the reflection spectrum in the rest-frame of the gas. In this model, the electron density is assumed to be a constant ($10^{15}$ cm$^{-3}$) across the entire disk. The ionization has instead a radial profile described by a power law
\begin{equation}
\xi(r)=\xi_\mathrm{in}\left(\frac{R_\mathrm{in}}r\right)^{\alpha_\xi}.
\end{equation}

For the index of $\alpha_\xi = 0$, model \texttt{relxillion\_nk} is reduced to \texttt{relxill\_nk}. Our best-fit result from model B4 indicates a positive value of $\alpha_\xi$, which shows the $\Delta\chi^2>80$ compared with Model B1, implying that the ionization parameter decreases as the radial coordinate $r$ increases within the actual disk.

\subsection{The Best-fit Model}\label{the best-fit model}

Among the initial models I1-I4 with deferent \texttt{relxill} flavors, Model I2$^+$ shows the best-fit results modeled by a high density disk, approximately 10$^{18}$ cm$^{-3}$. 
From the residuals of Model B1-B4 shown in Figure \ref{fig:model}, there are no significant differences among their fits. However, there is a slight elevation in the residuals above 30 keV among Models B1-B4. We attribute this behavior to a low count rate above 30 keV during the soft intermediate state.
If we compare their $\chi^2$, we see that model B4 best fits MAXI J1803-298, and its value of the ionization index is near 0.5. However, comparing the minimum of $\chi^2$ of different models is not a particularly robust method to determine which model is favored by the data. AIC (Akaike Information Criterion) \citep{akaike1974new}, which is already applied and discussed in \citet{mall2022impact}, is a more reliable method to determine the best model in this case of a relatively small size of samples for the number of free parameters. As we already have the minimum $\chi^2$ for every model by \texttt{XSPEC}, it can be calculated the AICc straightly by
\begin{equation}
\mathrm{AICc}=\chi_{\min}^2+2N_p+\frac{2N_p\left(N_p+1\right)}{\left(N_b-N_p-1\right)},
\end{equation}
where $N_p$ is the number of free parameters, and $N_b$ is the number
of bins. The values of AICc obtained from Model A1-A3 and Model B1-B4 are shown in Table \ref{tab:AICc}. As a general criterion for AIC, models with $\Delta AICc > 5$ are considered less favored by the data, while those with $\Delta AICc > 10$ are deemed ruled out and can be excluded from further analysis \citep{burnham2004model}. 

By applying this selection criterion, we determine that among Models A1-A3, Model I2$^+$, which incorporates a high-density disk, achieves the best fit. Among Models B1-B4, the \texttt{relxillion\_nk} model distinctly outperforms the other three in terms of fit quality. 
Despite we found a precise constraint on $\alpha_{13}=0.002^{+0.020}_{-0.015}$ with \texttt{relxillCP\_nk}, at a 3-$\sigma$ confidence level, Model B2 is deemed unsuitable for consideration. This is due to its fixed corona electron temperature at 100 keV and unrealistic estimates of certain parameters, such as $A_{Fe}$ and $log\xi$. Notably, the fits where $kT_e$ is allowed to vary, it invariably maxes out at 400 keV. Consequently, the outcomes from Models A2 and B2 are excluded from our analysis, not merely because of their elevated reduced $\chi^2$ values but also due to the low disk density values inserted. Considering the AICc values presented in Table \ref{tab:AICc} and their plausible parameter estimates, Models B3 and B4 are affirmed as providing more reliable constraints on $\alpha_{13}$.

% Consequently, a more reliable constraint on the deformation parameter $\alpha_{13}$ is determined as $\alpha_{13}=0.033^{+0.062}_{-0.033}$ , in 3-$\sigma$ credible lever if we consider the best-fit model.

\begin{longrotatetable}
\begin{deluxetable*}{ccccccccccc}
\tablenum{3}
\tablecaption{Summary of the Best-fit Values for Johannsen Models A1-B6 with only set free $\alpha_{13}$ after the MCMC Runs\label{tab:Parameters}}
\tablewidth{500pt}
\tabletypesize{\scriptsize}
\tablehead{
\colhead{} & \colhead{Model A1} & \colhead{Model B1} & \colhead{Model A2}  & \colhead{Model B2} & \colhead{Model A3}  & \colhead{Model B3}  & \colhead{Model A4} & \colhead{Model B4} & \colhead{Model B5} & \colhead{Model B6}
}
\startdata
{\texttt{tbabs}} & {}  & {}& {} & {}  & {}& {} & {} & {} & {}\\
{$N_H(10^{21} cm^{-2}$)}  & {$3.35^{+0.04}_{-0.08}$}& {$3.34^{+0.04}_{-0.03}$} & {$3.22^{+0.04}_{-0.03}$} & {$3.20^{+0.02}_{-0.03}$} & {$3.33^{+0.06}_{-0.05}$}& {$3.33^{+0.05}_{-0.04}$} & {$3.60^{+0.05}_{-0.06}$} & {$3.58^{+0.08}_{-0.06}$} & {$3.50^{+0.03}_{-0.06}$} & {$3.28^{+0.03}_{-0.03}$}\\
\hline
{\texttt{diskbb}} & {} & {} & {} & {} & {}& {} & {} & {} & {} & {}\\
{$T_{in}(keV)$}  & {$0.765^{+0.006}_{-0.015}$} & {$0.783^{+0.003}_{-0.003}$} & {$0.746^{+0.004}_{-0.003}$} & {$0.745^{+0.004}_{-0.004}$}  & {$0.774^{+0.005}_{-0.005}$}& {$0.777^{+0.006}_{-0.007}$} & {$0.808^{+0.007}_{-0.008}$} & {$0.804^{+0.012}_{-0.008}$}& {$0.787^{+0.004}_{-0.003}$} & {$0.755^{+0.002}_{-0.004}$}\\
{$N_{diskbb}$}  & {$606^{+53}_{-20}$} & {$599^{+14}_{-25}$} & {$702^{+14}_{-20}$} & {$714^{+15}_{-18}$}  & {$564^{+27}_{-22}$}& {$567^{+16}_{-41}$} & {$430^{+31}_{-24}$} & {$438^{+36}_{-37}$}& {$496^{+15}_{-18}$} & {$645^{+14}_{-9}$}\\
\hline
{\texttt{relxill\_nk flavor}} & {\texttt{relxill\_nk}} & {} & {\texttt{relxillCP\_nk}} & {}  & {\texttt{relxillD\_nk}}& {} & {\texttt{relxillion\_nk}} & {}& {\texttt{relxill\_nk}} & {}\\
{} & {\texttt{$\alpha_{13}=0$}} & {free $\alpha_{13}$} & {$\alpha_{13}=0$} & {free $\alpha_{13}$} & {$\alpha_{13}=0$}& {free $\alpha_{13}$} & {$\alpha_{13}=0$} & {free $\alpha_{13}$} & {\texttt{set Fe=1}} & {\texttt{set Fe=5}}\\
{$q_{in}$} & {$10^*$} & {$10^*$} & {$9.76^{}_{-0.92}$} & {$9.73^{}_{-0.91}$}  & {$9.94^{}_{-0.49}$}& {$9.94^{}_{-2.3}$} & {$9.73^{}_{-0.89}$} & {$9.96^{}_{-1.10}$}& {$9.94^{}_{-2.42}$} & {$9.94^{}_{-2.26}$}\\
{$q_{out}$} & {$1.62^{+0.75}_{-0.38}$} & {$1.83^{+0.44}_{-0.93}$} & {$2.02^{+0.26}_{-0.18}$} & {$2.09^{+0.27}_{-0.25}$}  & {$1.60^{+0.30}_{-0.28}$}& {$1.66^{+0.35}_{-0.56}$} & {$1.24^{+0.35}_{-0.35}$} & {$1.06^{+0.32}_{-0.55}$} & {$1.64^{+0.37}_{-0.77}$} & {$2.11^{+0.91}_{-1.21}$}\\
{$R_r$(M)} & {$2.97^{+0.47}_{-0.25}$} & {$2.85^{+1.36}_{-0.07}$} & {$2.83^{+0.15}_{-0.11}$} & {$2.80^{+0.33}_{-0.15}$}  & {$3.06^{+0.42}_{-0.20}$}& {$3.36^{+0.93}_{-0.36}$} & {$3.97^{+0.43}_{-0.49}$} & {$4.04^{+1.09}_{-0.42}$} & {$2.97^{+1.27}_{-0.33}$} & {$2.99^{+2.02}_{-0.13}$}\\
{$a^*$} & {$0.993^{}_{-0.011}$} & {$0.994^{}_{-0.012}$} & {$0.995^{+0.002}_{-0.005}$} & {$0.997^{+0.001}_{-0.002}$}  & {$0.992^{}_{-0.005}$}& {$0.992^{}_{-0.010}$} & {$0.988^{+0.004}_{-0.004}$} & {$0.993^{}_{-0.007}$} & {$0.997^{}_{-0.001}$} & {$0.995^{}_{-0.009}$}\\
{$i$(deg)} & {$71.1^{+3.0}_{-3.4}$} & {$73.7^{+1.2}_{-9.4}$} & {$69.2^{+2.8}_{-6.5}$} & {$69.7^{+3.1}_{-3.1}$}  & {$69.5^{+2.6}_{-5.4}$}& {$67.3^{+4.1}_{-9.8}$} & {$74.4^{+2.5}_{-2.4}$} & {$75.9^{+2.1}_{-3.3}$} & {$72.1^{+4.4}_{-3.9}$} & {$71.1^{+3.9}_{-6.1}$}\\
{$R_{in}$} & {$-1^*$} & {$-1^*$} & {$-1^*$} & {$-1^*$}  & {$-1^*$}& {$-1^*$} & {$-1^*$} & {$-1^*$} & {$-1^*$} & {$-1^*$}\\
{$\Gamma$} & {$2.203^{+0.013}_{-0.042}$} & {$2.204^{+0.023}_{-0.006}$} & {$2.187^{+0.013}_{-0.006}$} & {$2.178^{+0.008}_{-0.008}$}  & {$2.194^{+0.012}_{-0.011}$}& {$2.200^{+0.020}_{-0.009}$} & {$2.319^{+0.016}_{-0.021}$} & {$2.317^{+0.016}_{-0.020}$} & {$2.271^{+0.010}_{-0.011}$} & {$2.190^{+0.006}_{-0.008}$}\\
{$log\xi$} & {$4.29^{+0.25}_{-0.07}$} & {$4.30^{+0.02}_{-0.09}$} & {$4.84^{+0.10}_{-0.09}$} & {$4.80^{+0.12}_{-0.12}$}  & {$4.07^{+0.07}_{-0.06}$} & {$4.07^{+0.07}_{-0.05}$} & {$4.41^{+0.08}_{-0.07}$} & {$4.38^{+0.08}_{-0.08}$} & {$4.08^{+0.05}_{-0.03}$} & {$4.49^{+0.04}_{-0.04}$}\\
{$A_{Fe}$} & {$2.83^{+1.55}_{-0.50}$} & {$2.71^{+0.32}_{-0.47}$} & {$8.55^{+0.62}_{-0.94}$} & {$8.64^{+1.05}_{-0.98}$}  & {$2.02^{+0.50}_{-0.31}$}& {$1.89^{+0.61}_{-0.39}$} & {$1.41^{+0.46}_{-0.32}$} & {$1.31^{+0.71}_{-0.30}$} & {$1^*$} & {$5^*$}\\
{$E_{cut}(kT_e)(keV)$} & {$300^*$} & {$300^*$} & {$100^*$} & {$100^*$}  & {$300^*$}& {$300^*$} & {$300^*$} & {$300^*$} & {$300^*$} & {$300^*$}\\
{$logN(cm^{-3})$} & {$15^*$} & {$15^*$} & {$15^*$} & {$15^*$}  & {$18^*$}& {$18^*$} & {$15^*$} & {} & {$15^*$} & {$15^*$}\\
{$\alpha_{\xi}$} & {$0^*$} & {$0^*$} & {$0^*$} & {$0^*$}  & {$0^*$} & {$0^*$} & {$0.54^{+0.04}_{-0.04}$} & {$0.52^{+0.03}_{-0.04}$} & {$0^*$} & {$0^*$}\\
{$\alpha_{13}$} & {$0^*$} & {$0.029^{+0.043}_{-0.040}$} & {$0^*$} & {$0.002^{+0.021}_{-0.011}$}  & {$0^*$}& {$0.023^{+0.050}_{-0.031}$} & {$0^*$} & {$0.006^{+0.038}_{-0.016}$} & {$0.026^{+0.051}_{-0.050}$} & {$0.032^{+0.034}_{-0.037}$}\\
{$R_f$} & {$1.87^{+0.76}_{-0.64}$} & {$1.42^{+0.56}_{-0.20}$} & {$0.92^{+0.20}_{-0.17}$} & {$1.07^{+0.09}_{-0.16}$}  & {$1.64^{+0.14}_{-0.16}$}& {$1.03^{+0.40}_{-0.16}$} & {$0.96^{+0.12}_{-0.10}$} & {$0.98^{+0.30}_{-0.19}$} & {$1.47^{+0.69}_{-0.14}$} & {$1.25^{+0.46}_{-0.11}$}\\
{$norm(10^{-3})$} & {$3.9^{+2.1}_{-1.9}$} & {$4.1^{+2.3}_{-1.4}$} & {$7.0^{+1.8}_{-1.7}$} & {$6.2^{+1.8}_{-1.5}$}  & {$4.7^{+2.8}_{-1.7}$}& {$5.6^{+1.4}_{-0.8}$} & {$8.3^{+4.3}_{-4.6}$} & {$7.5^{+4.2}_{-3.7}$} & {$4.2^{+2.1}_{-3.2}$} & {$3.9^{+2.5}_{-3.1}$}\\
\hline
{\texttt{gaussian}} & {} & {} & {} & {}  & {} & {} & {} & {} & {} & {}\\
% {$E_{line}$} & {} & {} & {} & {$7.1^*$} & {} & {} & {}\\
{$\sigma_{line}$} & {$<0.31$} & {$<0.47$} & {$<0.13$} & {$<0.12$} & {$<0.33$} & {$<0.35$} & {$<0.53$} & {$<0.56$} & {$<0.41$} & {$<0.35$}\\
{$Norm(10^{-4})$} & {$-3.1$} & {$-2.6$} & {$-3.0$} & {$-2.3$}  & {$-4.1$}& {$-3.2$} & {$-3.1$} & {$-3.0$} & {$-2.6$} & {$-2.2$}\\
\hline
{$C_{FPMA}$} & {$1.023$} & {$1.023$} & {$1.023$} & {$1.023$}  & {$1.023$} & {$1.023$} & {$1.023$} & {$1.023$} & {$1.022$} & {$1.023$}\\
{$C_{FPMB}$} & {$0.996$} & {$0.997$} & {$0.997$} & {$0.995$}  & {$0.997$} & {$0.997$} & {$0.997$} & {$0.997$} & {$0.997$} & {$0.997$}\\
\hline
{\texttt{$\chi^{2}/\nu$(reduced)}} & {$2533/2281$} & {$2527/2280$} & {$2563/2280$} & {$2565/2279$} & {$2521/2280$}& {$2513/2279$} & {$2446/2279$} & {$2444/2278$} & {$2567/2280$} & {$2533/2280$}\\
 {} & {$=1.110$} & {$=1.108$} & {$=1.124$} & {$=1.125$} & {$=1.105$} & {$=1.103$} & {$=1.073$} & {$=1.072$}& {$=1.125$} & {$=1.111$}\\
\enddata
\tablecomments{Best-fit Values of Model A1-B6(includes \texttt{relxill\_nk}, \texttt{relxillCP\_nk}, \texttt{relxillD\_nk}, and \texttt{relxillion\_nk}). All errors determined through a 90\% confidence interval after MCMC runs for the global minimum. $*$ indicates that the parameter is frozen in the fit. $R_{in}=-1$ means that $R_{in}$ is set at the ISCO radius. The radial coordinate of the outer edge of the accretion disk is fixed to 400. $i$ is allowed to vary from $3^\circ$ to $80^\circ$. $a^*$ is allowed to vary from $-0.998$ to $0.998$. The deformation parameters of $\alpha_{13}$ is allowed to vary from $-1$ to 1. $E_{line}$ in \text{gaussian} is frozen to 7.1 keV. When the lower/upper uncertainty is not reported, the 90\% confidence level reaches the boundary (or the best-fit is at the boundary).}
\end{deluxetable*}
\end{longrotatetable}

\subsection{Constraints on $\alpha_{13}$ with MAXI J1803-298}\label{results on a13}

The constraints on the spin versus deformation parameter plane for Model B1-B4 after MCMC runs are shown in Figure \ref{fig:a13_a} ($a^*$ versus $\alpha_{13}$). The most precise measurements for $\alpha_{13}$ are measured as
\begin{equation}
\alpha_{13}=0.023^{+0.071}_{-0.038} ,
\end{equation}
from model B3 with \texttt{relxillD\_nk}, and 
\begin{equation}
\alpha_{13}=0.006^{+0.045}_{-0.022} ,
\end{equation}
from model B4 with \texttt{relxillion\_nk} in 3-$\sigma$ confidence level, respectively. These error bounds provided are purely statistical and do not include systematic errors.
%  where we need to note that 0.00 here is a very tiny value approach to 0.

The most stringent constraint on $\alpha_{13}$ to date has been obtained from
the analysis of a NuSTAR observation of the black hole in GX 339–
4 in \citet{tripathi2021testing} and its 3-$\sigma$ measurement of Johannsen deformation parameter on $\alpha_{13}$ is
\begin{equation}
\alpha_{13}=-0.02^{+0.03}_{-0.14} .
\end{equation}

We attribute the precision in determining $\alpha_{13}$ in this investigation is notably improved by the outstanding energy resolution and extensive effective area of the NICER detector, which excels in capturing reflection features.  Furthermore, our fitting process is likely unaffected by jet activities in BHXBs. Leveraging time-dependent visibility modeling of a relativistic jet, \citet{Wood2023MNRASmodellingjets} deduced the ejection date for MAXI J1803-298 as MJD 59348.08, predicting a cessation on May 23, 2021, in their radio flux density light curve. This timing corresponds with the dataset analyzed in our study, suggesting a significant reduction in jet activities as the system transitioned to the soft state.  We argue that such a pause in jet activities could be potentially one of the key factors in enhancing the precision of our $\alpha_{13}$ constraints.
% \textcolor{red}{Also mention that these error bounds are purely statistical. it does not include the systematic errors}

\subsection{Iron Abundance}

\begin{figure}[ht!]
\centering

\begin{minipage}[b]{0.3\linewidth}
  \centering
  Model B1\\
  \includegraphics[width=\linewidth]{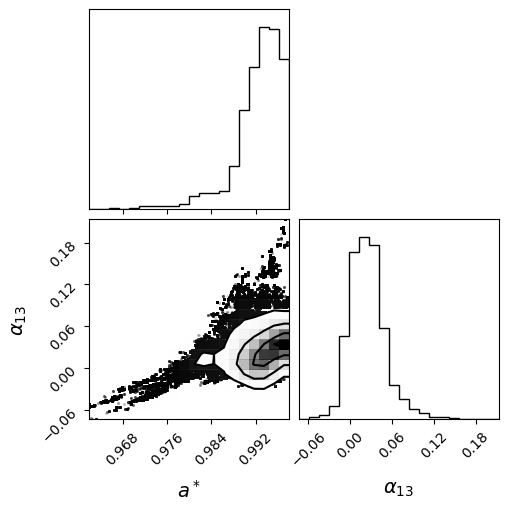}
\end{minipage}
% \hfill
\begin{minipage}[b]{0.3\linewidth}
  \centering
  Model B2\\
  \includegraphics[width=\linewidth]{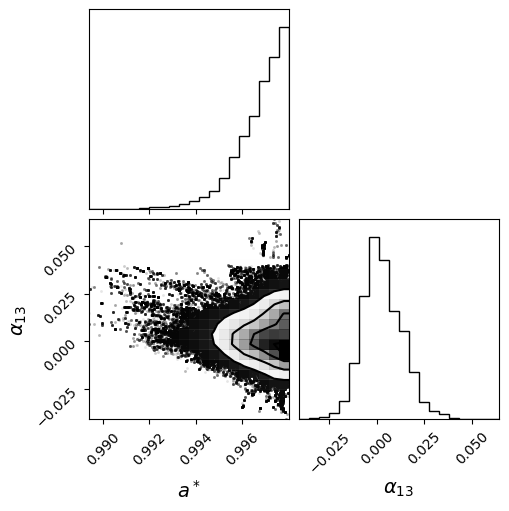}
\end{minipage}
\begin{minipage}[b]{0.3\linewidth}
  \centering
  Model B3\\
  \includegraphics[width=\linewidth]{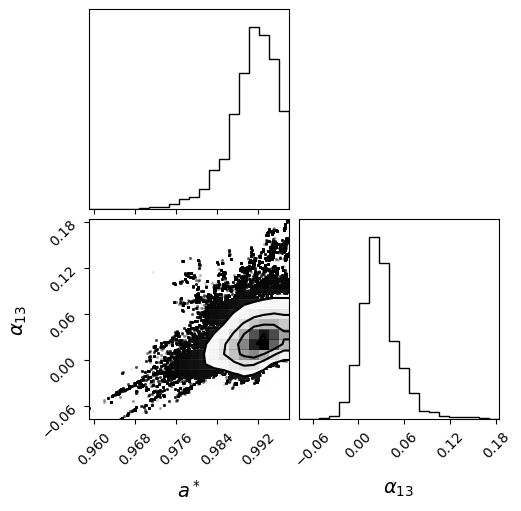}
\end{minipage}
% \hfill

\begin{minipage}[b]{0.3\linewidth}
  \centering
  Model B4\\
  \includegraphics[width=\linewidth]{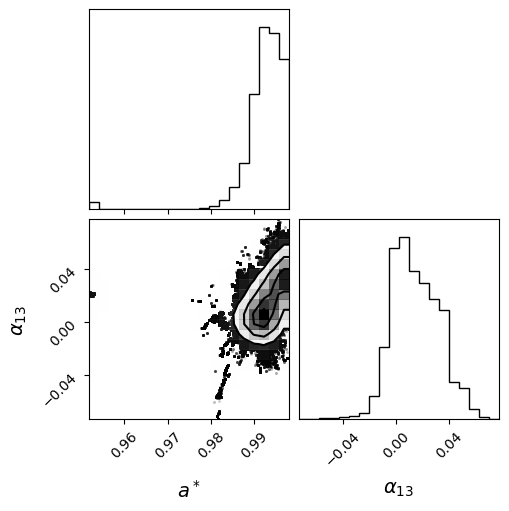}
\end{minipage}
\begin{minipage}[b]{0.3\linewidth}
  \centering
  Model B5 (\texttt{set Fe=1})\\
  \includegraphics[width=\linewidth]{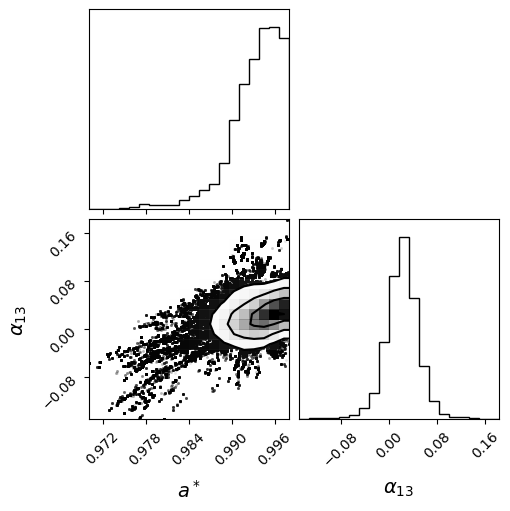}
\end{minipage}
\begin{minipage}[b]{0.3\linewidth}
  \centering
  Model B6 (\texttt{set Fe=5})\\
  \includegraphics[width=\linewidth]{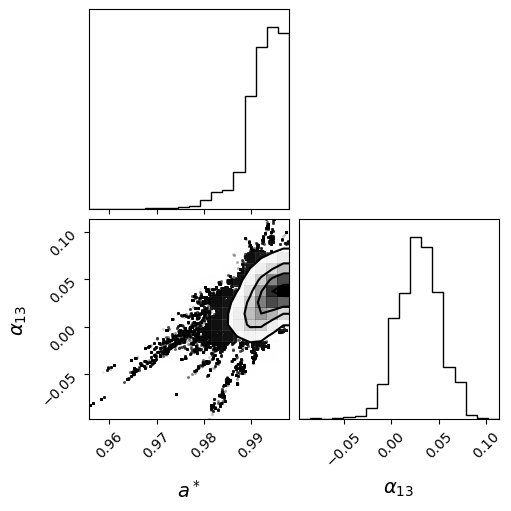}
\end{minipage}

\caption{Constraints on the spin parameter $a^*$ and the deformation parameter $\alpha_{13}$ for models B1-B6 after the MCMC runs. The 2D plots report the 1, 2, and 3$\sigma$ confidence contours.}
\label{fig:a13_a}
\end{figure}

The variable iron abundance ($A_{Fe}$) values in our employed models I1-I4 and A1-B4 appear to be inconclusive, that is because the reliability of iron abundance determinations derived from the analysis of accretion disk reflection spectra is a well-known issue \citep{bambi2021towards}. In particular, estimates of supersolar iron abundances become contentious in scenarios where reflection signatures are influenced by high plasma densities, as detailed by \citet{Garca2018ironproblem}, and the omission of returning radiation could lead to an underestimation of iron abundance \citep{riaz2021impact}. This concern is pertinent to our analysis of MAXI J1803-298. To mitigate potential biases in iron abundance estimates that could impact our constraints on $\alpha_{13}$, we incorporated Models B5 (with Fe=1) and B6 (with Fe=5) into our fitting process. Its constraints on the spin parameter versus $\alpha_{13}$ are shown in Figure \ref{fig:a13_a}, which shows no significant change to our results.

\section{The Concluding Remarks}\label{The Concluding Remarks}
It is noted that the reflection models we employed are simplified in current versions; see \citet{liu2019testing} for the list of simplifications in the current version of the model. In this present work, our model assumes an infinitesimally thin disk, a simplification, whereas actual disks possess finite thickness \citep{taylor2018exploring}. 
To ascertain whether the disk configuration of MAXI J1803-298 aligns with a geometrically thin disk model, \citet{Jana2022AstroSatJ1803} have estimated its accretion rate to be within the range of 6-10\% $L_{Edd}$ based on the ionization parameter, which means it is nicely in the 5\%–30\% range required by a Novikov–Thorne disk with an inner edge at ISCO radius \citep{Steiner2010isco}.
Despite our choice to position the inner edge of the accretion disk at the ISCO radius during a soft intermediate state, there might still be a discrepancy between our assumption and the actual disk configuration.
The ionization parameter $\xi$ is presumed to be either constant or subject to power-law decline across the disk. However, in reality, it is anticipated to exhibit a complex radial variation influenced by factors such as the X-ray flux from the corona and the disk density \citep{ingram2019public, kammoun2019steep}.

In this study, we employ a simplified broken power-law model to characterize the disk-corona geometry, recognizing that this represents a rudimentary approximation. In this approach, the calculation of the emissivity profile is omitted, while a consistent calculation based on the specific geometry between the corona and disk is reported by \citet{dauser2013irradiation}. Future work could also consider the impact of returning radiation in this work to obtain a more precise parameter space \citep{Mirzaev:2024fgd}. The launch of next-generation X-ray observatories in the coming years, such as Athena \citep{nandra2013hot} and eXTP \citep{zhang2019enhanced}, have the potential to significantly enhance our understanding of accretion processes through their improved instrumentation. Athena will incorporate innovative microcalorimeter technology to provide unprecedented energy resolution around iron emission lines. Meanwhile, eXTP will expand our ability to probe accretion physics. The capabilities of these advanced missions offer the exciting possibility of placing tighter constraints on models, thereby deepening our comprehension of these energetic phenomena.

\begin{deluxetable*}{ccccccc}
\tablenum{4}
\tablecaption{AICc values of the fits with I1-I4 and A1-B6 of MAXI J1803-298 in our study.\label{tab:AICc}}
\tablewidth{0pt}
\tablehead{
\colhead{Models} & \colhead{} & \colhead{} & \colhead{} & \colhead{} & \colhead{} & \colhead{}
}
% \decimalcolnumbers
\startdata
{\texttt{relxill flavor}} & I1 & I2 & I2$^+$ & I3 & I4& $-$ \\
{} & 2670.0 & 2666.4 & 2642.0 & {2592.3} & {2607.0} &{$-$} \\
\hline
{\texttt{relxill\_nk flavor} ($\alpha_{13}$ = 0)} & A1 & A2 & A3 & A4 & $-$ & $-$ \\
{} & 2633.0 & 2683.4 & 2642.4 & {2596.0} & {$-$} &{$-$} \\
\hline
{\texttt{relxill\_nk flavor} ($\alpha_{13}$ to free)} & B1 & B2 & B3 & B4 & B5 & B6 \\
{} & 2648.4 & 2715.0 & 2663.0 & {2634.0} & {2688.4} &{2654.4} \\
\enddata
% \tablecomments{}
\end{deluxetable*}

\acknowledgments

We thank the referee for constructive comments that helped us improve the quality of this paper. We thank Cosimo Bambi, Lijun Gou, and Yu Wang for constructive suggestions and fruitful discussions. This work was supported by the CAS `Light of West China' Program (grant No. 2021-XBQNXZ-005) and the National SKA Program of China (grant Nos. 2022SKA0120102 and 2020SKA0120300). M.G.N. acknowledges the support from the CAS Talent Program. L.C. acknowledges the support from the Tianshan Talent Training Program (grant No. 2023TSYCCX0099). M.G.N., A.T., and Y.F.H. acknowledge the support from the Xinjiang Tianchi Talent Program. Y.F.H. also acknowledges the support from the NSFC (grant No. 12233002) and the National Key R\&D Program of China (2021YFA0718500). This work was partly supported by the Urumqi Nanshan Astronomy and Deep Space Exploration Observation and Research Station of Xinjiang (XJYWZ2303).

\appendix

\section{Appendix information}
The Johannsen metric is a phenomenological deformation from the Kerr metric and is specifically designed for testing the Kerr black hole hypothesis with electromagnetic observations of black holes \citep{johannsen2013regular}. In BoyerLindquist-like coordinates, the line element is
\begin{equation}
\begin{aligned}
    ds^2 = & -\frac{\tilde{\Sigma}(\Delta-a^2 C_2^2 \sin^2 \theta)}{B^2} dt^2 + \frac{\tilde{\Sigma}}{\Delta C_5} dr^2 + \tilde{\Sigma} d\theta^2 - \frac{2a[(r^2+a^2)C_1 C_2 - \Delta] \tilde{\Sigma} \sin^2 \theta}{B^2} dt d\phi \\
    & + \frac{[(r^2+a^2)^2 C_1^2 - a^2 \Delta \sin^2 \theta] \tilde{\Sigma} \sin^2 \theta}{B^2} d\phi^2,
\end{aligned}
\end{equation}
where $M$ is the black hole mass, $a=J/M$ is the black hole angular momentum. $\tilde{\Sigma}=\Sigma+f$, and
\begin{equation}
\begin{aligned}
    \Sigma & =r^2+a^2 \cos ^2 \theta, \\
    \Delta & =r^2-2 M r+a^2, \\
    B & =\left(r^2+a^2\right) C_1-a^2 C_2 \sin ^2 \theta,
\end{aligned}
\end{equation}
in which the functions $f$, $C_1$, $C_2$, and $C_5$ are defined as 
\begin{equation}
\begin{aligned}
    f & =\sum_{n=3}^{\infty} \epsilon_n \frac{M^n}{r^{n-2}}, \\
    C_1 & =1+\sum_{n=3}^{\infty} \alpha_{1 n}\left(\frac{M}{r}\right)^n, \\
    C_2 & =1+\sum_{n=2}^{\infty} \alpha_{2 n}\left(\frac{M}{r}\right)^n, \\
    C_5 & =1+\sum_{n=2}^{\infty} \alpha_{5 n}\left(\frac{M}{r}\right)^n,
\end{aligned}
\end{equation}
where $\{\epsilon_n\}$, $\{\alpha_{1n}\}$,$\{\alpha_{2n}\}$, and $\{\alpha_{5n}\}$ are four infinite sets of deformation parameters without constraints from the Newtonian limit and solar system experiments. In this work, we have only considered the deformation parameter $\alpha_{13}$, which has the strongest impact on the reflection spectrum, and all other deformation parameters are assumed to vanish. To work with a regular metric, some constraints are imposed on the black hole spin parameter $a_*$ and the deformation parameter $\alpha_{13}$
\begin{equation}
-1 \leqslant a_* \leqslant 1, \quad \alpha_{13}>-\frac{1}{2}\left(1+\sqrt{1-a_*^2}\right)^4 ,
\end{equation}
as discussed in \citet{tripathi2018testing}.

\bibliography{sample63}{}
\bibliographystyle{aasjournal}

\end{document}